\documentclass[a4paper,11pt]{article}
\pdfoutput=1 % if your are submitting a pdflatex (i.e. if you have
\usepackage{jcappub} % for details on the use of the package, please
\usepackage{subcaption}
\usepackage{graphicx}
\usepackage{caption}
\usepackage{afterpage}
\usepackage{makecell}
\usepackage{bm}% bold math

\title{\boldmath Predictive uncertainty on improved astrophysics recovery from multifield cosmology}

\author[a,b,1]{Sambatra Andrianomena,\note{Corresponding author.}}
\author[c,b]{Sultan Hassan}
\affiliation[a]{South African Radio Astronomy Observatory (SARAO), Black River Park, Observatory, Cape Town, 7925, South Africa}
\affiliation[b]{University of the Western Cape, Bellville, Cape Town 7535,
South Africa}
\affiliation[c]{Flatiron Institute Center for Computational Astrophysics, 162 5th Ave 5th floor, New York, NY 10010, USA}

\emailAdd{andrianomena@gmail.com}
\abstract{We investigate how the constraints on cosmological and astrophysical parameters ($\Omega_{\rm m}$, $\sigma_{8}$, $A_{\rm SN1}$, $A_{\rm SN2}$) vary when exploiting information from multiple fields in cosmology. We make use of a convolutional neural network to retrieve the salient features from different combinations of field maps from IllustrisTNG in the CAMELS project. The fields considered are neutral hydrogen (HI), gas density (Mgas), magnetic fields (B) and gas metallicity (Z). We estimate the predictive uncertainty on the predictions of our model by using Monte Carlo dropout, a Bayesian approximation. Results show that overall, the performance of the model improves on all parameters as the number of channels of its input is increased. As compared to previous works, our model is able to predict the astrophysical parameters with up to 5\% higher in accuracy. In the best setup which includes all fields (four channel input, Mgas-HI-B-Z) the model achieves $R^{2} > 0.96$ on all parameters. Similarly, we find that the total uncertainty, which is dominated by the aleatoric uncertainty, decreases as more fields are used to train the model in general. 
The uncertainties obtained by dropout variational inference are overestimated on all parameters in our case, in that the predictive uncertainty is much larger than the actual squared error. After calibration, which consists of a simple $\sigma$ scaling method, the average deviation of the total uncertainty from the actual error goes down to $25\%$ at most (on $A_{\rm SN1}$).       
}

\begin{document}
\maketitle
\flushbottom

\section{Introduction}\label{sec:intro}
In order to retrieve the relevant information from observables (e.g. galaxy, cosmic shear) for predicting underlying cosmology, summary statistic, such as power spectrum, has been used extensively (e.g. \cite{ivanov2020cosmological}). The scale of interest in those studies is such that the influence of baryons is not important so that the theory (i.e. power spectrum) can still be leveraged to harvest the cosmological information. However inferring the cosmological parameters in  a highly non-linear regime, where baryonic physics can not be neglected, is a non-trivial task. \cite{jing2006influence} demonstrated the increasing effect of baryons on weak lensing power spectrum on small scales such that precision cosmology on those scales will need to account for those effects. \cite{levine2006active} suggested that an elaborate model of Active Galactic Nuclei (AGN) outflows is required to explain their effect on the matter power spectrum. 

Another approach that looks very promising is to directly build a mapping between the distribution of matter and the corresponding cosmology. \cite{ravanbakhsh2016estimating} and \cite{pan2020cosmological} made use of convolutional neural network (CNN) to predict $\Omega_{\rm m}$ and $\sigma_{8}$ from simulated 3D distribution of dark matter. They reported that their neural network achieved better accuracy in comparison with power spectral analyses.  
% \cite{auld2007fast} fast cosmological parameter inference using neural network ML using CMB power spectrum as input 
\cite{gupta2018non} and \cite{ribli2019improved} utilized the predictive power of CNN to extract cosmological information from weak lensing maps, provided the wealth of information the latter carry.    
\cite{fluri2019cosmological},  who also resorted to deep learning to constrain cosmology using weak lensing maps, investigated the small scale influence of baryons on the constraints which were found to be negatively impacted. 
% They also highlighted the improvement     
% \cite{fluri2022cosmological} deep learning with weak lensing including baryon effects.
HI intensity mapping has been proven to be a powerful cosmological probe with the advent of SKA experiment \cite{santos2015cosmology}, since mapping out the HI distribution traces the underlying density and large scale structure. Few examples of CNN's success in the context of HI maps are the following. Using HI 2D maps from {\sc SimFast21} \cite{santos2008cosmic,santos2010fast,hassan2016simulating}, a semi-numerical model to evolve reionization on cosmological scales, \cite{hassan2020constraining} trained two different network architectures to infer the cosmological and astrophysical parameters from the HI field. \cite{hassan2019identifying} and \cite{mangena2020constraining} used similar network architectures to identify reionization sources, and to recontruct the reionization history respectively.
For the first time, \cite{villaescusa2021multifield} extracted cosmological and astrophysical parameters from multifield 2D maps, consisting of a set of fields where each one corresponds to a channel at the input of a deep regressor. They showed that their architecture is capable of marginalizing over the contamination by the astrophysical effects at the field level to constrain both the cosmology and astrophysics with a relatively high accuracy.

Within the context of inferring the cosmological and astrophysical parameters directly from a field distribution (3D, 2D maps) using deep learning, uncertainty on the model predictions has not been carefully estimated, to the best of our knowledge, a least not in a Bayesian sense. The models that have been considered generate point estimates so far. This poses an issue in that it is challenging to assess what the model has learned, e.g. it is likely to give a \textit{seemingly good} estimate even though the input is drawn from an out-of-distribution sample. Whereas in the case of a probabilistic model, which produces predictive uncertainty on its predictions, erroneous predictions which correspond to high uncertainty can be safely ignored. 

We present in this work the amount of information gained from considering different sets of fields when constraining cosmology and astrophysics. We investigate how the performance of a neural network and the resulting predictive uncertainty on each parameter improve by leveraging the key features from 2D maps with multiple channels. The data used in this work and the network model we consider are presented in Section~\ref{sec:dataset-model}. Results are shown in Section~\ref{sec:results} where the effects of the set of fields used for the training on both the model performance and the total uncertainty for each parameters are detailed in \ref{sec:performance} and \ref{sec:uncertainty} respectively. The use of a scaling-based approach to mitigate the miscalibration of the uncertainties produced by the Bayesian network is discussed in \ref{sec:sigma-scaling}. We finally conclude in Section~\ref{sec:conclusion}.

\section{Data and model}\label{sec:dataset-model}
\begin{figure}[!h]
\centering
\includegraphics[width=0.85\textwidth]{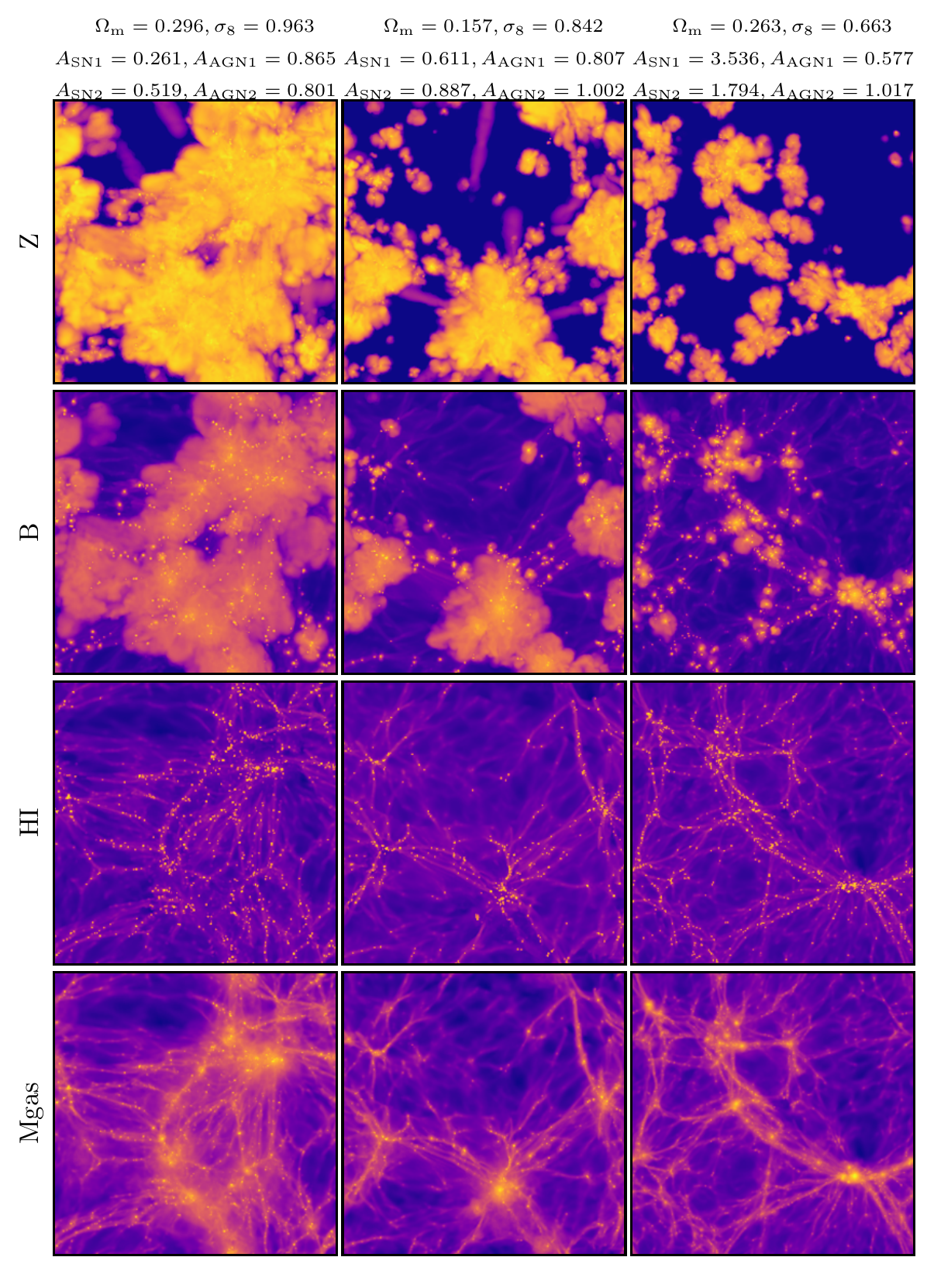}
\caption{Each map of size 25 $h^{-1}$Mpc is an example of each field that is considered in this study. Mgas (\textit{fourth row}), HI (\textit{third row}), B (\textit{second row}) and Z (\textit{first row}) indicate gas density maps, neutral hydrogen density maps, magnetic fields maps and gas metallicity maps respectively. The maps in the same column correspond to the same realization. The first column, second column and third column correspond to weak, modest and strong stellar feedback. 
% lower values  ($A_{\rm SN1}$ = 0.261, $A_{\rm SN2}$ = 0.519), middle values ($A_{\rm SN1}$ = 0.611, $A_{\rm SN2}$ = 0.887) and higher values ($A_{\rm SN1}$ = 3.536, $A_{\rm SN2}$ = 1.794) of the stellar feedback parameters respectively.
}
\label{fig:camels-field}
\end{figure}

The Cosmology and Astrophysics with MachinE Learning Simulations (CAMELS) project \cite{villaescusa2021camels, villaescusa2022camels, villaescusa2022camels1} consists of 4,233 numerical simulations which comprise dark matter only and state-of-the-art hydrodynamic simulations. The latter consider two different implementations of the subgrid physics models, namely,  IllustrisTNG \cite{nelson2019illustristng} and SIMBA \cite{dave2019simba}. CAMELS, whose main goal is to understand the theory that explains observables for a given cosmology and astrophysics, is specifically designed for training machine learning algorithms. The multifield dataset that we choose in this work is from IllustrisTNG simulation suite in the CAMELS project, and comprises 195000 2D maps as a whole, yielding 15000 maps for each of the thirteen fields (e.g. gas density, neutral hydrogen density). The maps were produced at $z = 0$ from varying cosmological ($\Omega_{m}$, $\sigma_{8}$), astrophysical parameters ($A_{\rm SN1}$, $A_{\rm SN2}$, $A_{\rm AGN1}$, $A_{\rm AGN2}$) and the initial seed number in thousands of simulation runs. The parameters $A_{\rm SN1}$ and $A_{\rm SN2}$ encode the supernova feedback, whereas $A_{\rm AGN1}$ and $A_{\rm AGN2}$ control the AGN feedback. For easy reference, we present in Table~\ref{tab:variation-range} the prior range of each parameter, and refer the interested reader to \cite{villaescusa2021camels} for more details. Each 2D map of 25 $h^{-1}{\rm Mpc}$ in size has $256\times256$ pixels resolution. 
\begin{table}[h!]
 \centering
 \begin{tabular}{|c|c|}
  \hline
   Parameter & Range \\
  \hline
  $\Omega_{m}$ & [0.1, 0.5] \\[2pt]
  $\sigma_{8}$ & [0.6, 1.0]\\[2pt]
   $A_{\rm SN1}$ & [0.25, 4]\\[2pt]
  $A_{\rm SN2}$ & [0.5, 2]\\[2pt]
  $A_{\rm AGN1}$ & [0.25, 4] \\[2pt]
  $A_{\rm AGN2}$ & [0.5, 2] \\[2pt]
  \hline
 \end{tabular}
 \caption{Prior range of each parameter.}
 \label{tab:variation-range}
\end{table}
For our analyses we only select four fields, namely gas density (Mgas), neutral hydrogen density (HI), magnetic field (B) and gas metallicity (Z). The fields have been selected based on how sensitive they are to the variation of the parameters of interest in this study (see \cite{villaescusa2021multifield}). Figure~\ref{fig:camels-field} shows the four fields considered in this study in three different realizations which are related to three different sets of cosmological and astrophysical parameters, with different varying feedback strengths. The four fields -- Z (first row), B (second row), HI (third row) and Mgas (fourth row) -- in the same column correspond to one realization. As this set of parameters are highly degenerate (e.g. low $A_{\rm SN2}$ and high $A_{\rm AGN2}$ would boost star formation in the outskirts of galaxies and push gas to low dense regions, see Figure 2 in \cite{villaescusa2021multifield}), it is challenging to determine the leading effect. However, the following combination of $\Omega_{m}=0.296, \sigma_{8}=0.963, A_{\rm SN1}=0.261, A_{\rm AGN1}=0.865, A_{\rm SN2}=0.519, A_{\rm AGN2}=0.801$ (left column) shows the scenario where gas is being pushed out to less dense regions. The right column shows the combination of $\Omega_{m}=0.263, \sigma_{8}=0.663, A_{\rm SN1}=3.536, A_{\rm AGN1}=0.577, A_{\rm SN2}=1.794, A_{\rm AGN2}=1.017$, where the gas is concentrated at the high dense regions. The middle column has a moderate feedback strength between the left and right columns with parameters $\Omega_{m}=0.157, \sigma_{8}=0.842, A_{\rm SN1}=0.611, A_{\rm AGN1}=0.807, A_{\rm SN2}=0.887, A_{\rm AGN2}=1.002$.

%For illustration, we select sets with low, middle and high values of the stellar feedback parameters ($A_{\rm SN1}$ and $A_{\rm SN2}$) for the first, second and third columns respectively.} 
% shows 2D maps generated with the same set of cosmological/astrophysical parameters and initial seed number. 

In order to build a mapping between the input (2D maps) and the underlying cosmology and astrophysics, we resort to a deep convolutional network, whose architecture, which is quite similar to VGG \cite{simonyan2014very} , is provided in Table~\ref{tab:architecture}. It is worth pointing that the number of channels of the inputs varies according to the types of setup which are presented in Table~\ref{tab:setups}. 
\begin{table}[h!]
 \centering
 \begin{tabular}{|c|c|c|}
  \hline
   & Layer & (in channel, out channel, kernel, stride) \\
  \hline
  1 & Batchnorm & ($n$, $n$, --, --) \\[2pt]
  2  & ConvConv& ($n$, 32, 3$\times$3 , 1)\\[2pt]
  3 & Maxpooling & (32, 32, 3$\times$3, 2)\\[2pt]
  4  & ConvConv & (32, 64, 3$\times$3 , 1)\\[2pt]
  5 & Maxpooling & (64, 64, 3$\times$3, 2) \\[2pt]
  6 & ConvConvConv & (64, 128, 3$\times$3 , 1)\\[2pt]
  7 & Maxpooling & (128, 128, 3$\times$3, 2)\\[2pt]
  8 & ConvConvConv & (128, 256, 3$\times$3 , 1)\\[2pt]
  9 & AdaptiveAveragepooling & (256, 256, --, 2)\\[2pt]
  10  & Flatten & --\\[2pt]
  11 & Dropout & --\\[2pt]
  12 & $\rm{Fully\ Connected\ Layer}$ &  (1024, 512, --, --)\\[2pt]
  13 & ReLU & --\\[2pt]
  14 & Dropout & --\\[2pt]
  15 & $\rm{Fully\ Connected\ Layer}$ &  (512, 12, --, --)\\[2pt]
  \hline
 \end{tabular}
 \caption{The architecture of the network in this work. The letter $n$ indicates the number of channels (or also maps) of each input, ConvConv and ConvConvConv denote chaining two and three\textit{
 Convolution+Batch Normalization+ReLU} layers respectively.}
 \label{tab:architecture}
\end{table}
To estimate the predictive uncertainty via Monte Carlo Dropout \cite{gal2016dropout} we have two dropout layers, each with probability of 0.2. The choice of the dropout rate is based on \cite{gal2016dropout}. The first 6 components from the network outputs are the mean of parameters to be constrained ($\Omega_{m}$, $\sigma_{8}$, $A_{\rm SN1}$, $A_{\rm SN2}$, $A_{\rm AGN1}$, $A_{\rm AGN2}$) which we denote by $\mu_{i}$ ($i = 1,...,6$) and the remaining components are the standard deviation for each parameter $\theta_{i}$, denoted by $\sigma_{i}$. This construction is such that the aleatoric uncertainty, which is the other component of the total uncertainty, can be estimated. Following the prescription in \cite{villaescusa2021multifield}, the optimization (using Adam optimizer) via mini-batch gradient descent is achieved by using the loss function
\begin{eqnarray}\label{loss-function}
    \mathcal{L} &=& \frac{1}{N_{\rm param}}\sum_{i = 1}^{N_{\rm param}}{\rm log}\left(\frac{1}{M_{\rm batch}}\sum_{j = 1}^{M_{\rm batch}}(\theta_{i,j} - \mu_{i,j})^{2}\right) \nonumber\\
   && + \frac{1}{N_{\rm param}}\sum_{i = 1}^{N_{\rm param}}{\rm log}\left(\frac{1}{M_{\rm batch}}\sum_{j = 1}^{M_{\rm batch}}\left((\theta_{i,j} - \mu_{i,j})^{2} - \sigma_{i,j}^{2}\right)^{2}\right),
\end{eqnarray}
where $N_{\rm param}$ and $M_{\rm batch}$ are the number of parameters (6) and batch size (64) respectively. The first term in Equation~\ref{loss-function} is the mean squared error whereas the second term enforces the aleatoric uncertainty to be consistent with the squared error such that it is well calibrated.       

\begin{table}[h!]
 \centering
 \begin{tabular}{|c|c|}
  \hline
   & Setup \\
  \hline
  HI & number of channels = 1, only the HI maps are fed into the network \\[2pt]
  Mgas-HI & number of channels = 2, stack of Mgas and HI maps as inputs\\[2pt]
  Mgas-HI-B & number of channels = 3, stack of Mgas, HI and B maps as inputs\\[2pt]
  Mgas-HI-B-Z & number of channels = 4, stack of Mgas, HI, B and Z maps as inputs\\[2pt]
  \hline
 \end{tabular}
 \caption{Different setups for the analyses.}
 \label{tab:setups}
\end{table}
We use $80\%$ of the dataset for training which is run for 200 epochs with batch size of 64 and the remaining instances are used for validation and testing. In total, there are four training runs for four different setups (see Table~\ref{tab:setups}).
\section{Results}\label{sec:results}
In this section, the performance of the model is assessed using coefficient of determination $R^{2}$, a metric that indicates how well the model is capable of explaining the proportion of variance in the predicted parameters. Following \cite{villaescusa2021multifield}, we also use the relative accuracy which is defined as $\langle \delta\theta_{i}/\theta_{i}\rangle$ where $\theta_{i}$ are the parameters to be constrained and $\delta\theta_{i} = |\theta^{\rm true}_{i} - \theta^{\rm predicted}_{i}|$.
\subsection{Performance}\label{sec:performance}
\begin{figure}[!h]
\centering
\includegraphics[width=0.8\textwidth]{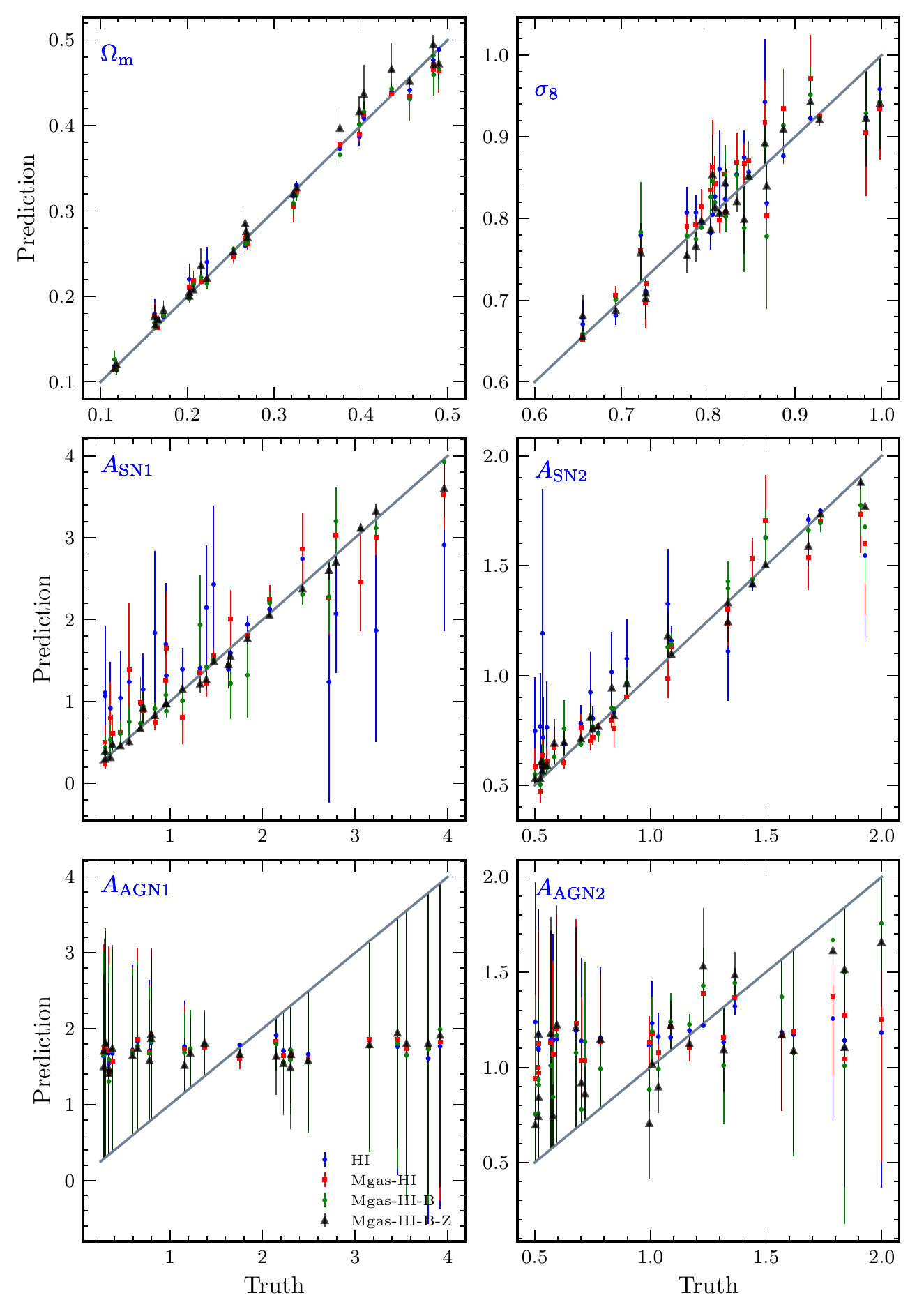}
\caption{Predictions for each parameter. Each type of marker corresponds to predictions in one setup (e.g. 2D maps of neutral hydrogen density as inputs). The errorbars indicate the absolute difference between the the target and network output ($\mu_{i}$). The network performs well on the cosmological ($\Omega_{\rm m}$, $\sigma_{8}$) and stellar feedback ($A_{\rm SN1}$ and $A_{\rm SN2}$) parameters with Mgas-HI-B-Z setup. However constraining the AGN feedback parameters is challenging.}
\label{fig:predictions_all_params}
\end{figure}

We show in Figure~\ref{fig:predictions_all_params} the predictions for all 6 parameters in each setup. The errorbars denote the absolute difference between ground truth and prediction, and the solid grey line is the identity line. For better visualization, the plots in Figure~\ref{fig:predictions_all_params} are obtained by selecting random examples from the test set and their corresponding predictions by the network. It can be noticed how the performance of the model differs between setups, which are indicated by different markers. 

% \begin{figure}[!h]
% \centering
% \includegraphics[width=1\textwidth]{prediction_truth_all_params.pdf}
% \caption{Predictions for each parameter. Each type of marker corresponds to predictions in one setup (e.g. 2D maps of neutral hydrogen density as inputs).}
% \label{fig:predictions_all_params}
% \end{figure}
Results suggest that inferring $A_{\rm AGN1}$ and $A_{\rm AGN2}$ are challenging for the model, as indicated by the predictions, with larger errors, fluctuating around the same value in all setups. We find that $R^{2}$ values are -0.05 for $A_{\rm AGN1}$ and 0.50 for $A_{\rm AGN2}$, indicative of weak correlation (or no correlation at all) between their actual values and predictions. This trend is consistent with what was found by \cite{villaescusa2021multifield} in which the challenge related to these two parameters is attributed to either the maps are not very sensitive to them or the model complexity is such that it fails to retrieve the relevant information. For the rest of our analyses, we restrict ourselves to the four cosmological and astrophysical parameters of interest, i.e. $\Omega_{\rm m}$, $\sigma_{8}$, $A_{\rm SN1}$ and $A_{\rm SN2}$.

\begin{figure}[tbh]
\centering
\includegraphics[width=.9\textwidth]{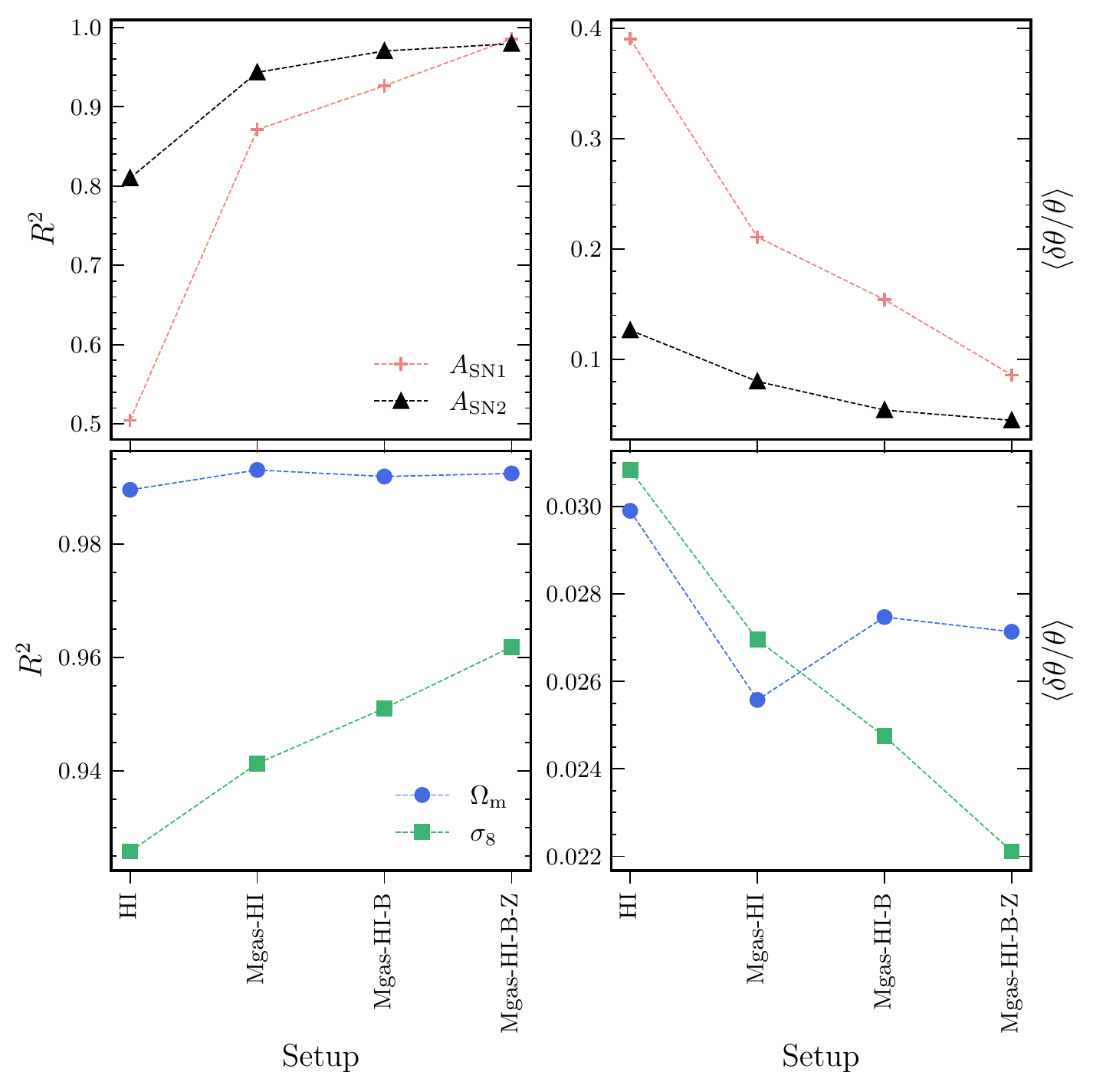}
\caption{Performance of the model. Panels on the left column show the coefficient of determination for each parameter as a function of the type of input. Panel on the right column present the variation of the average relative accuracy for each parameter as function of the setup. The cosmological and astrophysical parameters are plotted separately for better visualization. It is clear that extracting the cosmology and astrophysics improves with more information (more fields are stacked) at the input.}
\label{fig:metrics}
\end{figure}
We present in Figure~\ref{fig:metrics} the variation of $R^{2}$ and that of accuracy as a function of the type inputs on the left and right columns  respectively for the parameters ($\Omega_{\rm m}$, $\sigma_{8}$, $A_{\rm SN1}$, $A_{\rm SN2}$). For each metric, we plot the cosmology and astrophysics in different panels to clearly show how the metrics vary with the setups. Overall, it is clear that the more information is fed into the model the better its performance is. This is evidenced by the increasing $R^{2}$ value (see Figure~\ref{fig:metrics} left column) with an increasing number of the input channels. The cosmological parameters, especially the matter density, appear to be less sensitive to the number of fields (or channels), whereas the model performance on the astrophysical parameters greatly improves with more fields stacked at the model input. The $R^{2}$ value related to $A_{\rm SN1}$ goes from 0.5, when only HI maps are used as inputs, to 0.98 with the Mgas-HI-B-Z setup. The variation of accuracy as a function of the setup is consistent with that of the coefficient of determination. In other words, the predictions are more accurate when more information is provided to the model. It appears that the model accuracy on $\Omega_{\rm m}$ is best with the Mgas-HI setup, corresponding to the lowest relative error. But overall it's a small fluctuation, provided that the difference between the best and worst accuracy is $\sim 0.4\%$, which is insignificant. We argue that the overall increase in performance of the regressor has begun to reach a plateau with the Mgas-HI-B-Z setup ($R^{2}>$ 0.96 for all parameters). It is tempting to quote that the model only requires four different field maps from the CAMELS Multifield Dataset (CMD) to accurately determine the underlying cosmology and astrophysics. However, it is a bit premature to make such claim at this stage, given that we haven't tested yet all possible combinations between the eleven maps in CMD (excluding both dark matter density and dark matter velocity), which is beyond the scope of this work. \cite{villaescusa2021multifield} investigated the performance of their network when combining the 11 fields for the training. By comparing the accuracy of our network prediction with only four fields/channel inputs -- $\langle \delta\theta/\theta\rangle[\%]$ = 2.2, 2.7,  4.5, 8.6 for $\sigma_{8}$, $\Omega_{\rm m}$, $A_{\rm SN2}$, $A_{\rm SN1}$ respectively -- with theirs which was obtained from stacking eleven fields at the input -- $\langle \delta\theta/\theta\rangle[\%]$ = 2.3, 2.5,  4.8, 13.0 for $\sigma_{8}$, $\Omega_{\rm m}$, $A_{\rm SN2}$, $A_{\rm SN1}$ respectively -- the results are promising. Compared to \cite{villaescusa2021multifield}, our model is able to predict the astrophysical parameters with up to 5\% higher in accuracy. This could be either due to the difference in the network architectures used in each work or inferring the astrophysics might favor less fields as an input. It is entirely possible that more complex architectures might be needed to extract more information for large number of fields (i.e. 11). We leave investigating this effect to a follow-up work.

\begin{table}[h!]
 \centering
 \begin{tabular}{|l|c|c|}
  \hline
   \diaghead{\theadfont Diag ColumnmnHead II}%
{\normalsize Test set}{\normalsize Training set} & IllustrisTNG & SIMBA \\
  \hline
  IllustrisTNG & \makecell{$\Omega_{\rm m}$: 0.99 (0.027) \\ $\sigma_{8}$: 0.97 (0.018)} & \makecell{$\Omega_{\rm m}$: 0.99 (0.035) \\ $\sigma_{8}$: 0.96 (0.023)}\\[2pt]
  \hline
  SIMBA & \makecell{$\Omega_{\rm m}$: 0.97 (0.045) \\ $\sigma_{8}$: 0.92 (0.029)} & \makecell{$\Omega_{\rm m}$: 0.99 (0.028) \\ $\sigma_{8}$: 0.96 (0.019)}\\[2pt]
  \hline
 \end{tabular}
 \caption{Comparing the performance of our model on out-of-distribution sample with its performance on in-distribution sample. Values in each cell are $R^{2}$ and the accuracy $\langle\delta\theta/\theta\rangle$ (in round brackets).}
 \label{tab:generalization}
\end{table}

\cite{villaescusa2021robust} demonstrated how their model, trained on maps of total matter density (Mtot) generated from one simulation, was capable of inferring the cosmological parameters ($\Omega_{\rm m}$, $\sigma_{8}$) of Mtot maps from a completely different simulation to a relatively high accuracy. This demonstrates a good level of generalization of their network and points towards good model predictions on out-of-distribution sample in the context of cosmological inference. In order to understand the capacity of our network, we also train it on Mtot maps from one simulation suite in CAMELS to predict $\Omega_{\rm m}$ and $\sigma_{8}$ on Mtot maps from another simulation suite. Table~\ref{tab:generalization} summarizes our results. The network is perfectly capable of predicting the cosmological parameters from out-of-distribution maps, i.e. trained on IllustrisTNG maps and tested on SIMBA maps and vice versa. This is evidenced by both the relatively high values of $R^{2}$ ($>0.9$) and accuracy on all parameters for the two cases (IllustrisTNG - SIMBA and SIMBA - IllustrisTNG). Overall the values of the two metrics ($R^{2}$, $\langle\delta\theta/\theta\rangle$) in the case of out-of-distribution samples and in-distribution samples are comparable, which is indicative of a good generalization capability of our model in this specific setup, i.e. inferring cosmological parameters using Mtot maps.

\subsection{Uncertainty}\label{sec:uncertainty}
In a real world scenario, the \textit{frequentist} approach in deep learning is prone to overconfidence. Given its prediction which is point-wise estimate, assessing what the model is ignorant about (e.g. out-of-distribution instances) poses a challenge. Unlike \textit{Bayesian} deep learning in which prediction is associated with uncertainty, wrong prediction corresponds to large uncertainty, reflecting what the model doesn't know. In this work, we model the predictive uncertainty of our network  which results from the combination of the epistemic and aleatoric uncertainties. The former, also known as \textit{model uncertainty}, encodes the uncertainty in the network weights and can be reduced by having more data for the training. 
\begin{figure}[!h]
\centering
\includegraphics[width=1\textwidth]{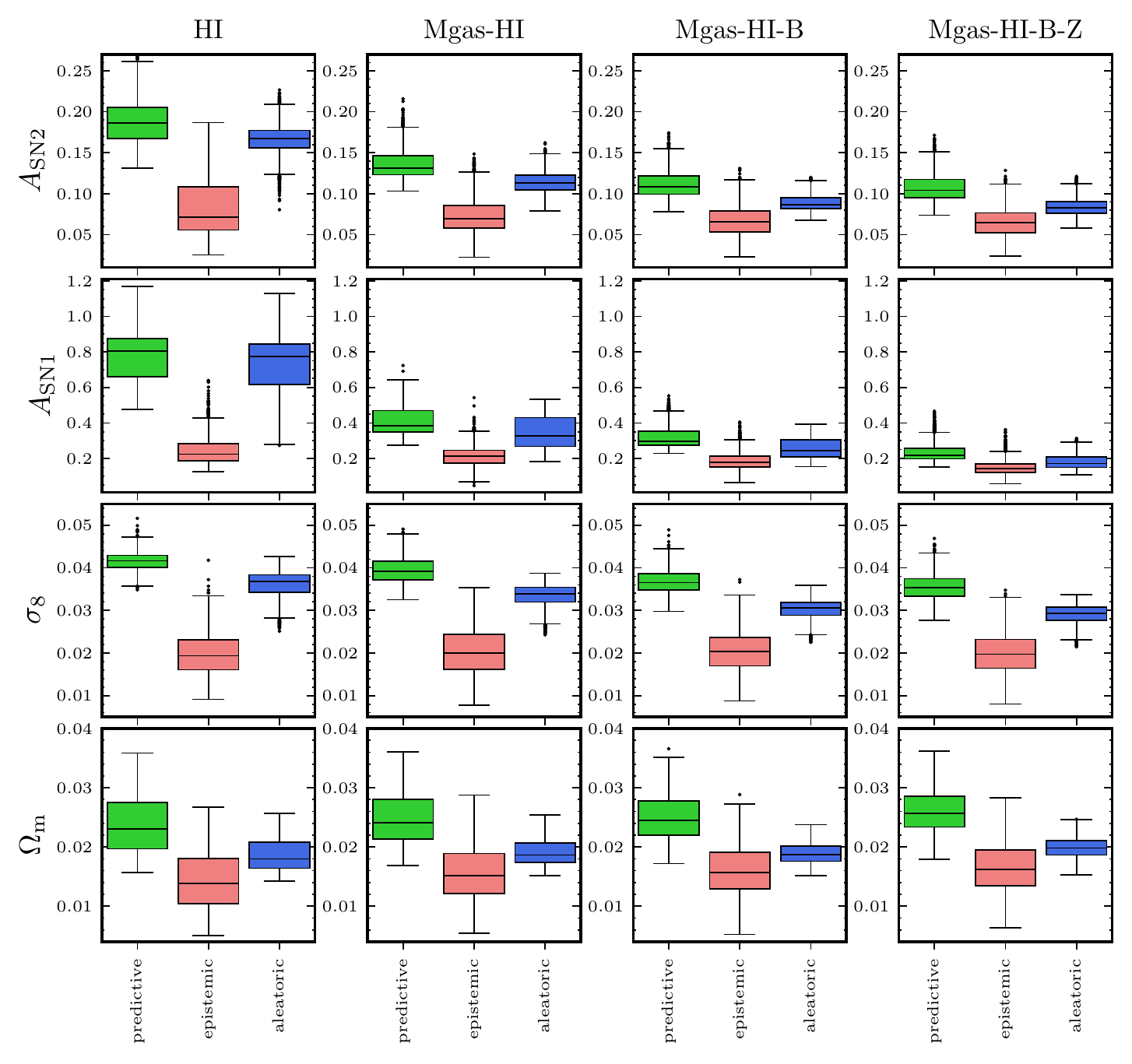}
\caption{Distributions of predictive, epistemic and aleatoric uncertainties for each parameter in each setup. The panels in the same column are associated with one setup, and those on the same row are the results for one parameter. Overall, it can be seen, especially for the astrophysical parameters, that the uncertainties improve with the increasing number of stacked fields.}
\label{fig:uncertainty_types}
\end{figure}
The latter accounts for the intrinsic noise in the data (e.g. resolution) and cannot be decreased by having more examples in the training set. 
We turn our frequentist model into a Bayesian model by enabling the dropout layers at test time 
% such that the epistemic uncertainty can be estimated by using samples from $T$ number of forward passes, which is set to 200 in our case.
 as in \cite{gal2016dropout} where the equivalence between neural network with dropout layers and an approximation to a Bayesian model like Gaussian process was demonstrated. We build our network such that its outputs comprise the mean $\hat{\bm{y}}$ and standard deviation $\hat{\bm{\sigma}}$, giving a total number of 12 outputs\footnote{We mentioned in the previous section that $A_{\rm AGN1}$ and $A_{\rm AGN2}$ would be left out for the rest of the work. However that does not change the number of the outputs of our network, which always includes all 6 parameters.}.
 \begin{figure}[!h]
\centering
\includegraphics[width=1\textwidth]{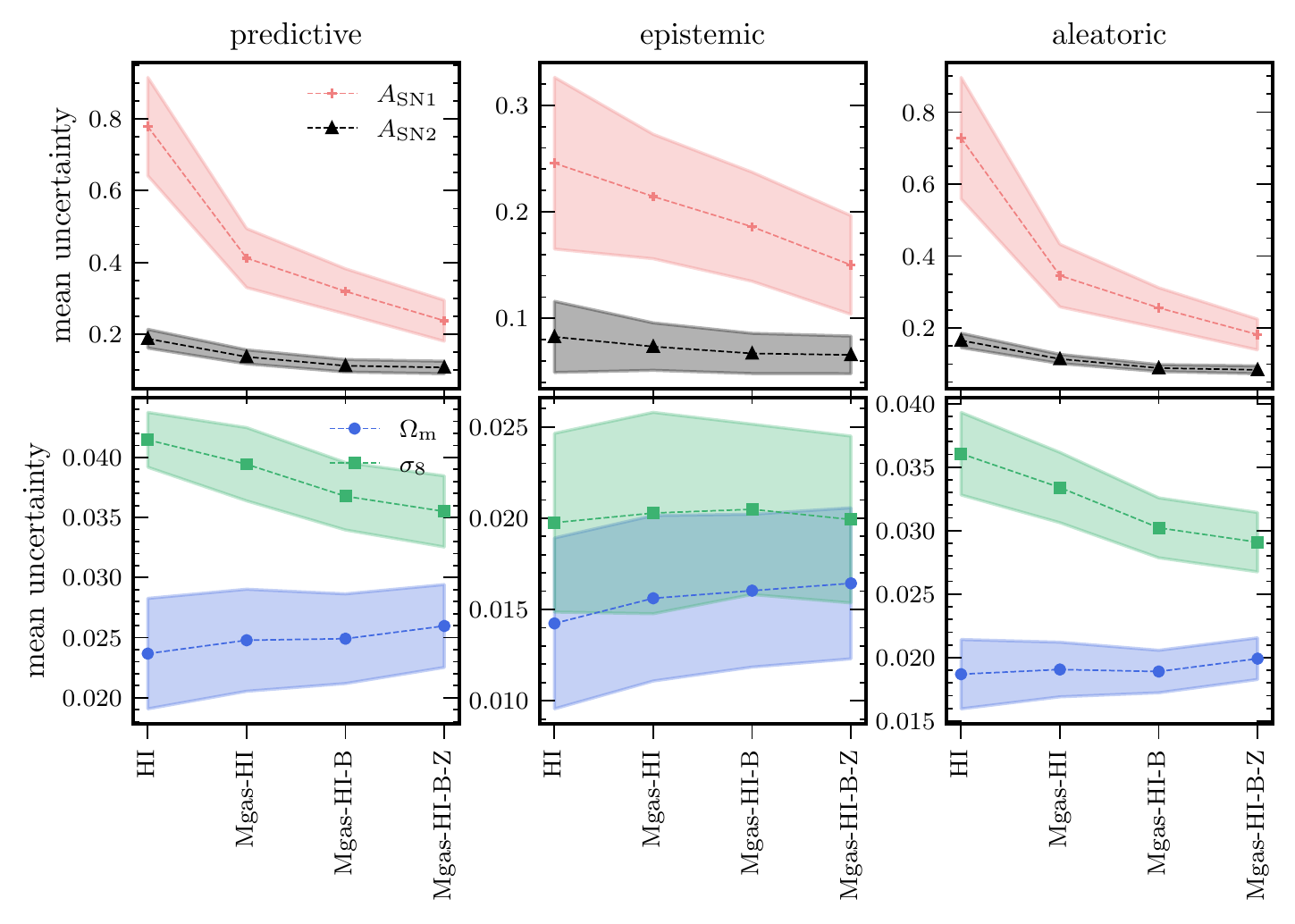}
\caption{Variation of the mean of each uncertainty distribution for each parameter as a function of the setup. The first and second rows show the mean uncertainties for astrophysical and cosmological parameters respectively. Markers indicate the mean value of a distribution and shaded area corresponds to the scatter of that distribution. The mean value of each type of uncertainty of each parameter decreases with more fields added, except for the case of $\Omega_{\rm m}$ where each type of uncertainty does not appear to be sensitive to the number of channels.}
\label{fig:median_uncertainties}
\end{figure}
 Following \cite{kendall2017uncertainties}, the predictive uncertainty (or also total uncertainty), resulting from adding the epistemic and aleatoric uncertainties in quadrature, is given by 
\begin{equation}\label{eq:uncertainty}
    \sigma_{\rm predictive}^{2} = \frac{1}{T}\sum_{t = 1}^{T}\hat{\bm{y}}_{t}^{2} - \left(\frac{1}{T}\sum_{t = 1}^{T}\hat{\bm{y}}_{t}\right)^{2} + \frac{1}{T}\sum_{t = 1}^{T}\hat{\bm{\sigma}}_{t}^{2},
\end{equation}
where $\hat{\bm{y}}_{t}$ and $\hat{\bm{\sigma}}_{t}$ are samples from $T$ number of forward passes which is set to 200 in our case. The first two terms in Equation~\ref{eq:uncertainty} denote the epistemic uncertainty, whereas the last term is the aleatoric uncertainty.

Figure~\ref{fig:uncertainty_types} shows the distributions of all types of uncertainty for each of the four parameters in each setup. Each colored box and their whiskers denote the interquartile range ($IQR$) and minimum/maximum respectively. The dots represent the outliers in the distribution. All panels in the same row show the distributions in each setup for one parameter. Based on the median value of the distribution, results suggest that the aleatoric uncertainty is a larger contributor to the uncertainty budget (total uncertainty). The predictive, epistemic and aleatoric uncertainties corresponding to the cosmological parameters ($\Omega_{\rm m}$, $\sigma_{8}$) are less sensitive to the amount of information provided to the network in that their change is marginal if any with the increasing number of channels of the input. However the positive effect of stacking more field maps for the training on the uncertainties is noticeable in the case of the astrophysical parameters ($A_{\rm SN1}$, $A_{\rm SN2}$). As the channel number is increased, both the \textit{IQR} and the median value of each type of distribution get smaller for the astrophysical parameters. A more compact representation of the distribution of each type of uncertainty as a function of the setup is shown in Figure~\ref{fig:median_uncertainties}. The latter presents the mean uncertainty, indicated by a marker, and the dispersion of uncertainty, denoted by shaded area, of each parameter in each setup. The cosmological and astrophysical parameters are plotted separately.  
% The variation of the median value of each type of distribution as a function of setup is more apparent in Figure~\ref{fig:median_uncertainties}. 
\begin{table}[h!]
 \centering
 \begin{tabular}{|c|c|}
  \hline
   Parameter & relative change in the mean of total uncertainty $[\%]$ \\
  \hline
  $\Omega_{m}$ & 9.72 \\[2pt]
  $\sigma_{8}$ & -14.37\\[2pt]
   $A_{\rm SN1}$ & -69.50\\[2pt]
  $A_{\rm SN2}$ & -42.77\\[2pt]
  \hline
 \end{tabular}
 \caption{Relative change in mean total uncertainty between one channel input (HI) to four channel input. Negative value means decrease in uncertainty, in other words tighter constraint on average, and positive value means weaker constraint on average.}
 \label{table:relative-change}
\end{table}
The constraints get tighter with more fields used to train the model and the improvement is more significant for the astrophysical parameters, especially $A_{\rm SN1}$. The mean predictive uncertainties (and also the mean aleatoric ones) related to the matter density appear to slightly increase with the number of channels at the model input. However, by taking the variability of the uncertainty into account, that increase is only marginal. Provided the dynamic range of each parameter, we further investigate the change in the mean value of predictive uncertainty ($\bar{\sigma}_{\rm pred}^{\rm setup}$) by computing its relative increase/decrease from only using HI maps to using Mgas-HI-B-Z maps to train the model. Table~\ref{table:relative-change} presents the relative change in mean of predictive uncertainty  according to
\begin{equation}
    \Delta\overline{\sigma}_{\rm predictive} = 100 \times \frac{\overline{\sigma}_{\rm predictive}^{\rm{Mgas-HI-B-Z}} - \overline{\sigma}_{\rm predictive}^{\rm HI}}{\overline{\sigma}_{\rm predictive}^{\rm HI}}.
\end{equation}
%The weaker constraint on $\Omega_{\rm m}$ obtained with Mgas-HI-B-Z setup, as evidenced by the increase in $\Delta\sigma_{predictive}$ (positive), is consistent with the decrease in relative accuracy, i.e. larger value of $\langle\delta\theta/\theta\rangle$, in Figure~\ref{fig:metrics}. 
% The weaker constraint on $\Omega_{\rm m}$ obtained with Mgas-HI-B-Z setup can be accounted for by the fact that the neutral hydrogen, a biased tracer of the underlying dark matter, is enough to constrain the matter density such that the inherent noise in the maps of other fields which do not bring more relevant information to predict $\Omega_{\rm m}$, slightly worsen the constraint on $\Omega_{\rm m }$.
The relative change in the mean predictive uncertainty is relatively small for the cosmological parameters. The constraints on the matter density parameter, with a corresponding $\Delta\overline{\sigma}_{\rm predictive} = 9.72\%$, seems marginally tighter with HI setup than with Mgas-HI-B-Z (Table~\ref{table:relative-change}) on average, however the overall dispersion of the predictive uncertainty (see Figure~\ref{fig:median_uncertainties} bottom left panel) shows that resulting total uncertainty on $\Omega_{\rm m}$ prediction is not really sensitive to the selected setup for the training. 
% however the  \textit{IQR} of the predictive uncertainty is smaller (see Figure~\ref{fig:uncertainty_types}) when combining the four fields. This points to the fact that the estimated uncertainties for $\Omega_{\rm m}$ in the case of Mgas-HI-B-Z are slightly more robust. 
The constraints on $\sigma_{8}$, with a corresponding $\Delta\overline{\sigma}_{\rm predictive} = -14.37\%$, slightly improve on average with Mgas-HI-B-Z setup. Unlike with the case of $\Omega_{\rm m}$, the downward trend (along with the channel number) of the predictive uncertainty on $\sigma_{8}$ is a bit more noticeable. Similar to the performance of the model when including all four fields for the training, the generated uncertainties on the cosmology seem to be less impacted by the information that the gas density, magnetic field and gas metallicity carry. However, as indicated in Table~\ref{table:relative-change}, the model has gained relevant extra information from the three other channels to significantly tighten the constraints on $A_{\rm SN2}$ and $A_{\rm SN1}$. The relatively large improvement in the constraints on the astrophysical parameters that control the stellar feedback ($A_{\rm SN2}$, $A_{\rm SN1}$) can be attributed mainly to the information added by the gas metallicity to which they are related. As a consistency check, we compute the relative change in the mean predictive uncertainty between using gas metallicity only as input and the Mgas-HI-B-Z setup and find an improvement by only $10.33\%$ and $7.59\%$ for the constraints on $A_{\rm SN1}$ and $A_{\rm SN2}$ on average respectively. 
% It was shown in \cite{villaescusa2021multifield} that the accuracy of their model gets better by about $38.4\%$ and $37.5\%$ on $A_{\rm SN1}$ and $A_{\rm SN2}$ respectively when 11 channel input was considered. 
% Figure~\ref{fig:median_uncertainties} shows the median values of uncertainties as a function of setup for all parameters. 

\subsection{Calibrating uncertainty}\label{sec:sigma-scaling}
Predictive uncertainty via variational inference with MC Dropout is prone to miscalibration \cite{laves2020well}, causing it to be understimated/overestimated. In our analyses, we assess how well calibrated the uncertainties produced by the model are and use a simple approach to mitigate the calibration issue. In what follows, we further restrict ourselves to the Mgas-HI-B-Z setup, as it corresponds to the best constraints for all four parameters overall, but the same approach can be done for the other setups. 

\cite{laves2020well} prescribed the expected uncertainty calibration error (UCE) to evaluate the miscalibration of uncertainty for regression tasks in deep learning. 
\begin{figure}[!h]
\centering
\includegraphics[width=1\textwidth]{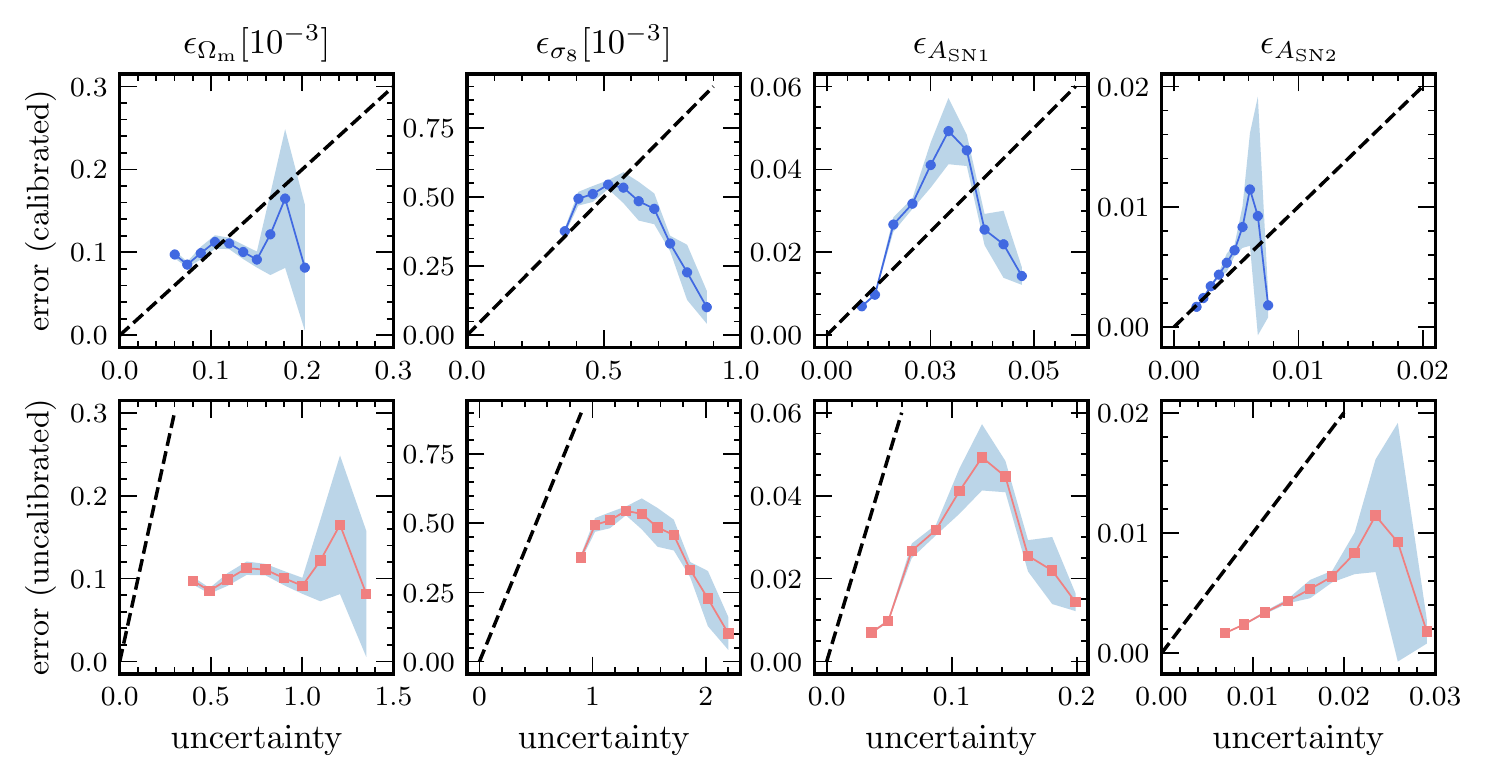}
\caption{Calibration of the model predictive uncertainty for $\Omega$, $\sigma_{8}$, $A_{SN1}$ and $A_{SN2}$, considering the combination of four different 2D maps as input, i.e. Mgas-HI-B-Z setup. The values related to the cosmological parameters are in $10^{-3}$ for better visualization. The uncertainties are overestimated, and applying $\sigma$ scaling method greatly improves the uncertainty calibration.}
\label{fig:calibration}
\end{figure}
The errors (between predictions and actual values) and their corresponding uncertainties from a test set are binned and the mean value of each quantity within a given bin is computed. UCE is the weighted average of the absolute difference between the mean error and mean uncertainty in all bins and given by  
\begin{equation}\label{eq:uce}
    {\rm UCE}[\%] = \sum_{m = 1}^{M}\frac{|B_{m}|}{N}|{\rm err}(B_{m}) - {\rm uncert}(B_{m})|,
\end{equation}
where $N$ is the total number of instances in the test set, $M$ indicates the number of bins, $B_{m}$ denotes the set of instances with a given bin, ${\rm err}(B_{m})$ the error in each bin given by
\begin{equation}\label{eq:average-error}
    {\rm err}(B_{m}) = \frac{1}{|B_{m}|}\sum_{i \in B_{m}} ||\hat{\bm{y}}_{i} - \bm{y}_{i}||^{2},
\end{equation}
and the average uncertainty in each bin ${\rm uncert}(B_{m})$ is 
\begin{equation}
    {\rm uncert}(B_{m}) = \frac{1}{|B_{m}|}\sum_{i \in B_{m}} (\sigma_{\rm predictive})_{i}.
\end{equation}
A perfect calibration is when UCE = $0\%$, i.e. the error against the uncertainty plot lies on the identity line.
% Provided the order of magnitude difference between parameters, a slightly different metric compared to prescrition in \cite{laves2020well} is used. 
We consider the weighted average of the relative difference between the uncertainty and the error, such that 
\begin{equation}\label{eq:uce-modified}
    {\rm UCE}_{\rm modified}[\%] = \frac{1}{M}\sum_{m = 1}^{M}\frac{|B_{m}|}{N}\left|\frac{{\rm err}(B_{m}) - {\rm uncert}(B_{m})}{{\rm err}(B_{m})}\right|.
\end{equation}
In other words, the relative deviation of the uncertainty with respect to the error in each bin is computed and the metric which is used to measure the miscalibration is the weighted average of that relative deviation. The main idea behind using ${\rm UCE}_{\rm modified}$ in Equation~\ref{eq:uce-modified} is to be able to compare the miscalibration related to each parameter.

\begin{table}[h!]
 \centering
 \begin{tabular}{|c|c|c|}
  \hline
   Parameter & ${\rm UCE}_{\rm modified}[\%]$, uncalibrated & ${\rm UCE}_{\rm modified}[\%]$, calibrated \\
  \hline
  $\Omega_{m}$ & 575.04 & 18.25\\[2pt]
  $\sigma_{8}$ & 155.90 & 15.81\\[2pt]
   $A_{\rm SN1}$ & 348.02 & 25.87\\[2pt]
  $A_{\rm SN2}$ & 250.42 & 12.81\\[2pt]
  \hline
 \end{tabular}
 \caption{Uncertainty calibration}
 \label{tab:uce-modify}
\end{table}
Bottom row in Figure~\ref{fig:calibration} shows the calibration plots for all parameters for uncalibrated uncertainty. On the same test set, we run five inferences, each with 200 forward passes, to investigate the variation of the uncertainties in each bin. Each red dot represents the mean of average error 
\begin{equation}\label{eq:mean-avg-err}
    \overline{{\rm err}(B_{m})} = \frac{1}{R}\sum_{r = 1}^{R}\{{\rm err}(B_{m})\}_{r}
\end{equation}
as a function the mean of the average uncertainty in a bin
\begin{equation}\label{eq:mean-avg-unc}
    \overline{{\rm uncert}(B_{m})} = \frac{1}{R}\sum_{r = 1}^{R}\{{\rm uncert}(B_{m})\}_{r},
\end{equation}
where $R$ is the number of runs. The blue shaded area denotes the variation in each bin and the black dashed line is the identity line. It is clear that the predictive uncertainty is poorly calibrated, as indicated by the values of ${\rm UCE}_{\rm modified}$ in the second column of Table~\ref{eq:uce-modified} which are all well above $100\%$. It is overestimated as it is larger than the actual error in prediction in all bins.\\
There are various methods for calibrating uncertainty, 
% \cite{kuleshov2018accurate, laves2020well, levi2019evaluating} 
ranging from involved ones such as building a mapping between the uncertainty and the error by using a multilayer perceptron \cite{kuleshov2018accurate} to simpler scaling based approach \cite{laves2020well, levi2019evaluating}. As a simple illustration of mitigating miscalibration of uncertainties generated by our Bayesian network, 
we opt for $\sigma$ scaling \cite{laves2020well} which is an approach that consists of computing a scalar $\bm{\mathcal{S}}$ that scales down\footnote{This could also be scaling up if the uncertainty is underestimated.} the uncertainties. For each parameter, we have that \cite{laves2020well}

\begin{equation}
    \bm{\mathcal{S}} = \sqrt{\frac{1}{N_{\rm valid}}\sum_{i = 1}^{N_{\rm valid}}\frac{||\hat{\bm{y}}_{i} - \bm{y}_{i}||^{2}}{(\sigma_{\rm predictive}^{2})_{i}}},
\end{equation}
where $N_{\rm valid}$ is the number of instances in the validation set which
we use and find that $\bm{\mathcal{S}}$ = 0.38, 0.63, 0.48, 0.51 for $\Omega_{\rm m}$, $\sigma_{8}$, $A_{\rm SN1}$, $A_{\rm SN2}$ respectively. For calibration, the uncertainties produced by the model on the test set are then multiplied by $\bm{\mathcal{S}}$. Top row in Figure~\ref{fig:calibration} shows the calibrated uncertainties using $\sigma$ scaling. The resulting ${\rm UCE}_{\rm modified}$ is given in the third column of Table~\ref{tab:uce-modify}. Results shows that after calibration, the predictive uncertainty deviates on average by $\sim 25\%$ (on $A_{\rm SN1}$) from the actual error at most and its smallest deviation is about $12\%$ (on $A_{\rm SN2}$).

\section{Conclusion}\label{sec:conclusion}
We have demonstrated in this work the exploitation of information from combining different fields to better understand the underlying cosmology and astrophysics by using Bayesian network. To this end, we have trained a deep convolutional neural network to retrieve the salient features from 2D multifield maps in order to constrain the cosmological ($\Omega_{\rm m}$, $\sigma_{8}$) and astrophysical ($A_{\rm SN1}$, $A_{\rm SN2}$) parameters. We have considered neutral hydrogen (HI), gas surface density (Mgas), magnetic field (B) and gas metallicity (Z) maps generated by the IllustrisTNG simulation suite in the CAMELS project. To investigate how the amount of information extracted from the fields impacts on both the performance of the model and the predictive uncertainty it produces, we have selected four different setups. In four separate training runs, we have selected: HI maps (HI), stacking Mgas and HI maps (Mgas-HI), stacking Mgas, HI and B maps (Mgas-HI-B), and stacking Mgas, HI, B and Z maps (Mgas-HI-B-Z) as inputs to the network. We have modeled the aleatoric uncertainty by having a mean ($\mu_{i}$) and standard deviation ($\sigma_{i}$) for each parameter at the output of our network. The total uncertainty is estimated by using MC dropout method, running 200 stochastic forward passes at inference time. 
% The predictive uncertainty (total uncertainty) produced by our network is obtained by combining the epistemic and aleatoric uncertainties. 
We have also assessed the miscalibration of the uncertainties generated by our network and used a simple scaling based approach to mitigate it. 

Overall, the network is capable of retrieving the salient features from the maps to predict the cosmological parameters and the astrophysical ones ($A_{\rm SN1}$, $A_{\rm SN2}$) which control the stellar feedback. The two parameters that encode the AGN feedback, however, pose a challenge for the model. This can be accounted for by either the model complexity or the fact that the fields don't contain enough information for the mapping. 
The regressor only requires the amount of information contained in maps of neutral hydrogen to determine the matter density parameter to a good level of accuracy, as indicated by an $R^{2} \sim 0.98$ and accuracy $\langle \delta\theta/\theta\rangle \sim 2.9$. This is expected since neutral hydrogen is a biased tracer of the underlying dark matter. Our finding comes to corroborate the results obtained in \cite{villaescusa2021multifield}. 
By increasing the number of channels (or fields) at the input of the network, the latter only marginally improves its performance on predicting $\Omega_{\rm m}$, achieving $R^{2} \sim 0.99$ and
$\langle\delta\theta/\theta\rangle \sim 2.7\%$. 
The results however suggest that the performance of the network on predicting $\sigma_{8}$, $A_{\rm SN1}$ and $A_{\rm SN2}$ increases to varying degrees as more fields are considered for the training. The network gains great amount of information from including the magnetic fields and gas metallicity in the input so as to constrain $A_{\rm SN1}$ and $A_{\rm SN2}$, as was already shown by \cite{villaescusa2021multifield}. By combining all the fields considered in this work, the performance of the model on all four parameters ($\Omega_{\rm m}$, $\sigma_{8}$, $A_{\rm SN1}$ and $A_{\rm SN2}$) is the best, as evidenced by $R^{2}>0.96$. Our model, with four channels at its input, has improved the accuracy on the astrophysical parameter $A_{\rm SN1}$ by $5\%$ compared to previous work \cite{villaescusa2021multifield} where eleven fields were stacked at the input of their model.

To assess the capacity of our network, we do predictions on out-of-distribution sample and compare them with predictions on in-distribution test set. Using the same set of hyperparameters for the main setups in this work, we train the model with the total matter density Mtot maps from IllustrisTNG and infer $\Omega_{\rm m}$ and $\sigma_{8}$ on test sets from IllustrisTNG and SIMBA separately, and vice versa. Results show that the performance of the network on in-distribution sample is only marginally better than its performance on out-of-distribution sample. This points towards the fact that our model is able to attain a promising level of generalization, identifying the relevant features from the total matter density maps to predict the cosmology, with the amount of data used for training.

We find a larger contribution of the aleatoric uncertainty to the total uncertainty budget on average. In all setups, for all four parameters, the median (or also the mean) value of the aleatoric uncertainty distribution is higher than that of epistemic uncertainty. However, except with the case of $A_{\rm SN1}$, the \textit{IQR} of the latter is broader than the former in general. It is interesting to see that the epistemic uncertainty (encoding the model ignorance), which can be mitigated by using more data,  decreases on average for the astrophysics by combining more fields, which does not augment the number of examples for the training but only provides more extra key features to the network. The decrease is more pronounced for $A_{\rm SN1}$. This implies that, in our case, the effect of adding more channels is to help the model better understand regions of parameter space with few available data. This is the same effect of adding more data (more examples) in order to minimize the epistemic uncertainty in less explored regions. The aleatoric uncertainty also decreases but more rapidly with more channels for the parameters that encode the galactic winds; owing to the fact that the model is able to beat down the inherent noise (such as resolution) in the maps by exploiting the extra information each field provides. The constraints on $\Omega_{\rm m}$ appear to be insensitive to having more channels at the input of the model. 
% On the one hand, the total uncertainty on matter density parameter with the Mgas-HI-B-Z setup degrades by about $11\%$ on average with respect to that of the HI setup. 
% This can be explained by the astrophysical effects in other maps and the fact that each field added to HI doesn't carry extra features to tighten the constraint but however is still subject to inherent noise, hence the slight increase in the predictive uncertainty. 
However the constraints on the amplitude of density fluctuations, the galactic winds parameters $A_{\rm SN1}$ and $A_{\rm SN2}$ in the Mgas-HI-B-Z setup improve by $\sim 14\%$, $\sim 69\%$ and $\sim 42\%$ compared to those in the HI setup respectively.

The predictive uncertainties on all parameters are all poorly calibrated. Results show that on average the relative deviation of the total uncertainty ($\rm UCE_{modified}$) on all parameters are at least twice as big and can even reach $6\times$ larger than the squared error on the setup that corresponds to the best constraints on all four parameters, i.e. Mgas-HI-B-Z. This indicates that the predictive uncertainty is overestimated. To remedy this miscalibration of the uncertainty, we adopt a simple scaling based approach which consists of computing the overall ratio $\bm{\mathcal{S}}$ between the squared error and the total uncertainty using the validation set. Calibration amounts to multiplying the uncertainties generated by the model when predicting the parameters of the test set by $\bm{\mathcal{S}}$. After calibration, the average deviation of the predictive uncertainty from the squared error goes down to about $25\%$ at most (on $A_{\rm SN1}$). The best calibrated uncertainty, or the smallest $\rm UCE_{modified}$, corresponds to $A_{\rm SN2}$ ($12\%$). 

Within the context of parameter inference in multifield cosmology, the results in this work look promising, but there are limitations that need to be highlighted. The maps used to train our network are noiseless, in that the systematics from instruments and observations like in a realistic scenario are not considered, e.g. foreground noise in HI intensity mapping. It is quite interesting to see how the performance of the network and the uncertainties will be impacted by the systematics and we defer this for future work. The field are selected based on the amount of information they carry about the cosmological and astrophysical parameters, as demonstrated in \cite{villaescusa2021multifield}. The objective is to investigate the improvement on the constraints by combining maps, but we have only considered four setups from a few selected fields as a first step. For future work, based on the results obtained in this work, it is possible to search for the optimized combination(s) that provide the maximum amount of relevant information that the model can exploit to further tighten the constraints. And the question:``\textit{what minimum subset of these fields enables us to get a fraction of those constraints?"} \cite{villaescusa2022camels} will also be addressed.

% Using the total matter density maps from IllustrisTNG for the training, we also test the generalization capability of our network by comparing its performance on a test set from IllustrisTNG with its performance on a test set from SIMBA.    
% , each corresponding to a number of fields at the model input, to map a 2D projection of     
% the effect of having more information by combining more maps to train the network on both its performance and the predictive uncertainties on its predictions, in the context of parameter inference. 

% \appendix
% \section{Gaussian Process (GP)}\label{gaussian_process}
% Known as a non-parametric model, the main assumption in a Gaussian Process is 

\acknowledgments

SA acknowledges financial support from the {\it South African Radio Astronomy Observatory} (SARAO). SH acknowledges support from  Simons Foundation. The authors are thankful to Francisco Villaescusa-Navarro for the useful discussions.

\bibliographystyle{JHEP}
\bibliography{multifield}

\end{document}